\documentclass[twocolumn]{svjour2}
\usepackage{amsmath}
\usepackage{graphicx}

\def\etal{{et~al.}}
\def\ER{Erd\"os-R\'enyi}
\def\BA{Barab\'asi-Albert}

\begin{document}

\title{The role of caretakers in disease dynamics}

\author{Charleston Noble \and James P.~Bagrow \and Dirk Brockmann}

\institute{C. Noble
\at Northwestern University, Evanston, IL 60208, USA
\and
J.P. Bagrow \and D. Brockmann
\at Engineering Sciences and Applied Mathematics,\\ 
Northwestern Institute on Complex Systems, \\
Northwestern University, Evanston, IL 60208, USA\\
\email{brockmann@northwestern.edu}
}

\date{September 12, 2012}

\maketitle

\begin{abstract}
One of the key challenges in modeling the dynamics of contagion phenomena is to
understand how the structure of social interactions shapes the time course of a
disease. Complex network theory has provided significant advances in this
context. However, awareness of an epidemic in a population typically yields
behavioral changes that correspond to changes in the network structure on which
the disease evolves. This feedback mechanism has not been investigated in depth.
For example, one would intuitively expect susceptible individuals to avoid other
infecteds. However, doctors treating patients or parents tending sick children
may also \emph{increase} the amount of contact made with an infecteds, in an
effort to speed up recovery but also exposing themselves to higher risks of
infection.  We study the role of these \emph{caretaker} links in an adaptive
network models where individuals react to a disease by increasing or decreasing
the amount of contact they make with infected individuals. We find that pure
avoidance, with only few caretaker links, is the best strategy for curtailing an
SIS disease in networks that possess a large topological variability.  In more
homogeneous networks, disease prevalence is decreased for low concentrations of
caretakers whereas a high prevalence emerges if caretaker concentration passes a
well defined critical value.
\end{abstract}

\section{Introduction}

Physicists have taken numerous approaches to modeling infectious diseases,
ranging from simple, deterministic compartmental models that qualitatively
describe disease dynamics in single populations~\cite{anderson:1992}, to highly
complex, stochastic metapopulation models that can account for the spread of
emergent infectious diseases on a global
scale~\cite{Ferguson:2005gp,Ferguson:2006p509,VandenBroeck:2011dj}.  Simple
models, designed to investigate the basic mechanisms underlying disease
dynamics, typically assume that a population is well-mixed, that interacting
individuals are identical and that stochastic effects are
negligible~\cite{ANDERSON:1979we,Brockmann:2010el}. On the other hand,  complex
computational models are manufactured to predict the time-course of actual
emergent infectious diseases such as H1N1 in 2009~\cite{Bajardi:2011gn}, SARS in
2003~\cite{Hufnagel:2004kt} quantitatively.  They typically take into account
data on social variability, age structure, spatial heterogeneity, seasonal
variation of disease dynamic parameters, multi-scale mobility networks, and
account for stochastic effects. Both classes of models fulfill equally
important, complementary, but almost mutually exclusive purposes.

Theoretical epidemiology experienced a major thrust with
the advent complex network theory and its introduction into the field~\cite{BarabasiAlbert2002,Newman2003}.
The study of network properties substantially advanced our understanding
of disease dynamic phenomena on multiple levels~\cite{ScaleFreeEpidemics}.
On one hand, networks were used as a model for inter-individual relationships
(social networks)~\cite{Newman:2002p963}. On the other hand, the network
approach was applied on a larger scale, modeling mobility
and transport between populations~\cite{Hufnagel:2004kt,Colizza:2007p521}.

The use of network theoretical concepts allowed researchers to investigate how
topological properties of underlying networks shape the contagion processes that
evolve on
them~\cite{ComputerViruses2001,SmallWorldEpidemic,SmallWorldStability,EpidemicThresholdScaleFree,RandomAssortiveScaleFree}.
In the context of epidemiology, mapping structural features of networks to
properties of the spread of the disease substantially increased the predictive
power of models and our understanding of epidemic phenomena.

Although it is intuitive and plausible that network features determine the
spread of a disease, it is equally plausible that an epidemic reshapes the
structure of the underlying networks. For example, in response to information on
an ongoing epidemic, people may change their behavior. They may decide to wear
face masks, avoid contacts, and travel less.  Surprisingly, this feedback
mechanism has been neglected even in some of the most detailed and sophisticated
modeling approaches~\cite{Ferguson:2005gp,Ferguson:2006p509}.  Topological
properties of social networks affect disease dynamics, and the disease then
feeds back to change the topology of the network. In order to understand the
dynamics of contagion phenomena in a population, it is vital to understand the
consequences of this feedback mechanism.

Networks that change their structure in response to their environment are called
\textit{adaptive}~\cite{GrossBlasius2007,BornholdtRohlf2000,HolmeGhoshal2006,Synchronization2011}.
In a recent study, Gross \etal{} proposed a simple adaptive network scheme,
based on a \textit{rewiring} rule, to understand how individuals' behavioral
changes impact on the time course of an epidemic. In this model, susceptible
individuals are allowed to protect themselves from infection by rewiring their
existing links~\cite{GrossLimaBlasius2006}. Specifically, with probability $w$ a
susceptible breaks the relationship with an infected person and forms a new link
to another, randomly selected susceptible. Despite the simplicity of this
approach, the mechanism can generate an abundance of interesting phenomena
including hysteresis and multi-stability.

Although this mechanism is attractive, the response to an ongoing epidemic in a
population has many facets. Not only do individuals avoid other infected
individuals (negative response). In many scenarios, individuals increase their
interaction with infected individuals (positive response), particularly in
hospital scenarios, and families in which individuals adopt the role of a
caretaker. Potentially, these positive responses can facilitate disease
proliferation in a population and yield a higher disease prevalence.  However,
caretaker activity can have a positive effect on infected individuals, for
example by increasing a person's recovery rate. A key question is how these
effects interact and under what circumstances caretaker activity has a net
positive or negative effect and how these effects play out in different network
topologies.

Here we propose and investigate these questions using an adaptive network model.
We consider two types of networks. First, the generic \ER{} random network with
binomial degree distribution, where each pair of nodes is linked with constant
probability $p_{ER}$~\cite{ErdosRenyi1960,Newman2003}.  We also consider \BA{}
scale-free networks with power law degree distributions, which more closely
mimic the heterogeneity in social interactions.  Dynamics on scale-free networks
have a number of important properties.  For instance, they lack epidemic
thresholds and are immune to random immunization due to strong connectivity
fluctuations~\cite{ScaleFreeEpidemics,EndemicStatesComplexNets,Newman2003,BarabasiAlbertScaling1999,Immunization2002}.
Thus diseases on scale-free networks are difficult to avoid, and once they take
hold, they are difficult to eradicate.  We will show that in scale free
topologies the highest disease extinction probabilities occur in the total
absence of caretakers, a surprising result which suggests that caretaker
relationships (including doctor/patient relationships) should be minimized in
those systems. For \ER{} networks we observe a critical caretaker proportion
which minimizes disease severity and beyond which additional caretakers increase
disease prevalence.

\section{Model description}

We consider a network with a constant number of nodes $N$, representing
individuals in a population. Each node is either susceptible ($S$) or infected
($I$). We denote the state variable of node $i$ by $x_{i}=0$ or $x_{i}=1$,
corresponding to states $S$ or $I$, respectively.  A pair $(i,j)$ of nodes share
a weighted symmetric link $w_{ij}\geq 0$ representing their contact rate. Note
that in general these contact rates can have any real positive value, unlike
network models that are based on binary interactions. Susceptible nodes can
become infected, and infected nodes can then become susceptible again upon
recovery.  This is the well-studied SIS (susceptible-infected-susceptible) model
\cite{Allen20001}. We also consider the SIR (susceptible-infected-recovered)
model where infected individuals become immune to the disease upon recovery.
Each link is designated either caretaker ($C$) or regular ($R$), and the
fraction of $C$ links is denoted $p_{c}$. We denote this signature of a link by
$\sigma_{ij}=1$ if the link is a caretaker link and $\sigma_{ij}=-1$ if it is
regular. These two classes represent different ways of responding to an
epidemic. Caretaker relationships cause nodes to increase their contact
frequency $w_{ij}$ if an attached node is infected, while regular relationships
cause nodes to avoid each other (decreasing contact rates). At each time step a
susceptible $i$ can become infected by one of its infected neighbors with a
probability that increases with link weight. We assume that: 
\begin{equation}
p_{i}=1-\exp\left(-\alpha_{i}\tau\right)\label{eq:infection_probability}
\end{equation}
where $\tau$ is the propensity of disease transmission following
a contact, and $\alpha_{i}=\sum_{j}w_{ij}x_{j}$ is the susceptible's
contact rate with infecteds.

\begin{figure}[t!]
\centering
\includegraphics[width=0.75\columnwidth]{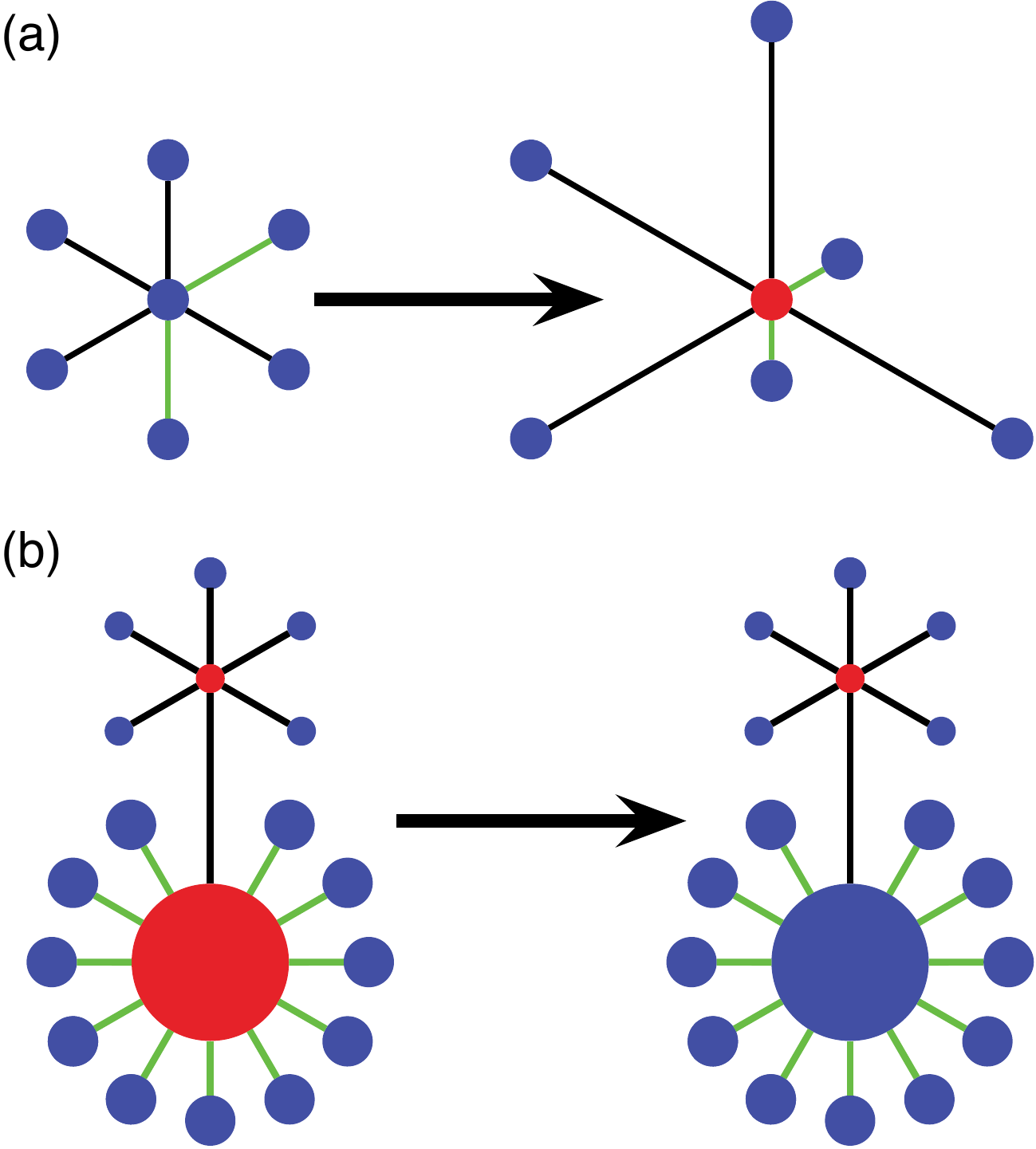}
\caption{(a) An initial network with all nodes susceptible (left) has two caretaker
links (green) and three regular links (black). After the infection
of the central node (shown by change to red color), regular-linked
nodes react by {}``avoiding'' the infected node (represented here
by increasing distance). Caretaker-linked nodes, on the other hand,
react by further increasing contact rates (represented here by decreasing
distance). (b) Another network consists of two clusters around two
central infected nodes (red). When considering the {}``caretaker
effect'', the more caretaker interactions (green) a node is exposed
to, the greater its recovery rate (shown by node size; larger nodes
have faster recovery rates). Thus after a time step, the lower infected
node is more likely to recover, shown by its transition to susceptible
status (blue). \label{fig:model_diagram}}
\end{figure}

An infected individual $i$ recovers with propensity $\beta_{i}$
which yields the probability of recovery 
\begin{equation}
r_{i}=1-\exp\left(-\beta_{i}\right)\label{eq:caretaker_effect}
\end{equation}
We consider two scenarios: 1) Infected nodes recover at a uniform
rate $\beta_{i}=\beta$ or 2) with variable probability. In the latter
case, caretaker relationships increase a node's recovery probability
$\beta_{i}$ according to \[
\beta_{i}=\beta_{0}+\left(\beta_{1}-\beta_{0}\right)\frac{\sigma_{i}^{n}}{\sigma_{0}^{n}+\sigma_{i}^{n}}\]
where $\beta_{0}$ is the base recovery rate, and $\beta_{1}$ the enhanced
recovery rate induced by the action of caretakers. The quantity $\sigma_{i}$
represents the total exposure of an infected to caretakers and is
given by \[
\sigma_{i}=\frac{1}{2}\sum_{j}w_{ij}(1+\sigma_{ij}),\]
thus $\sigma_{i}$ is the total weight of caretaker interactions
that node $i$ experiences. The parameter $\sigma_{0}$ sets the scale
for this exposure. The shape of the sigmoid curve can be controlled
by  the exponent $n$.

The infectious state of the system is defined by the states $x_{i}$ of each
node. We model the adaptive nature of the network weights $w_{ij}$ according to
\begin{equation}
    \delta_{t}w_{ij}=\mu\sigma_{ij}(x_{i}+x_{j})-\gamma\left(w_{ij}-w_{ij}^{0}\right).
    \label{eq:cont_time}
\end{equation}
Here the first term acts as the driving force of weight change, governed by the
rate parameter $\mu$. If a link is a caretaker link ($\sigma_{ij}=1)$, and one
of the adjacent nodes is infected ($x_{i}=1$ or $x_{j}=1$), this term is
positive and causes the weight to increase (if both nodes are infected the
change is additive). Regular links ($\sigma_{ij}=-1$), on the other hand
decrease in strength if one of the connected nodes is infected. The second term
acts as a restorative force, governed by the rate parameter $\gamma\ll\mu$.
Because we investigate a system in discrete time we use the following update
rule for the weights: 
\begin{equation}
    w_{ij}(t+1)=w_{ij}(t)\exp\left[\mu\sigma_{ij}(x_{i}+x_{j})-\gamma\left(w_{ij}(t)-w_{ij}^{0}\right)\right],
    \label{eq:discrete_time}
\end{equation}
a discrete time reformulation of Eq.~\eqref{eq:cont_time}.

\section{Results}

\begin{figure}[t!]
\centering
\includegraphics[width=\columnwidth]{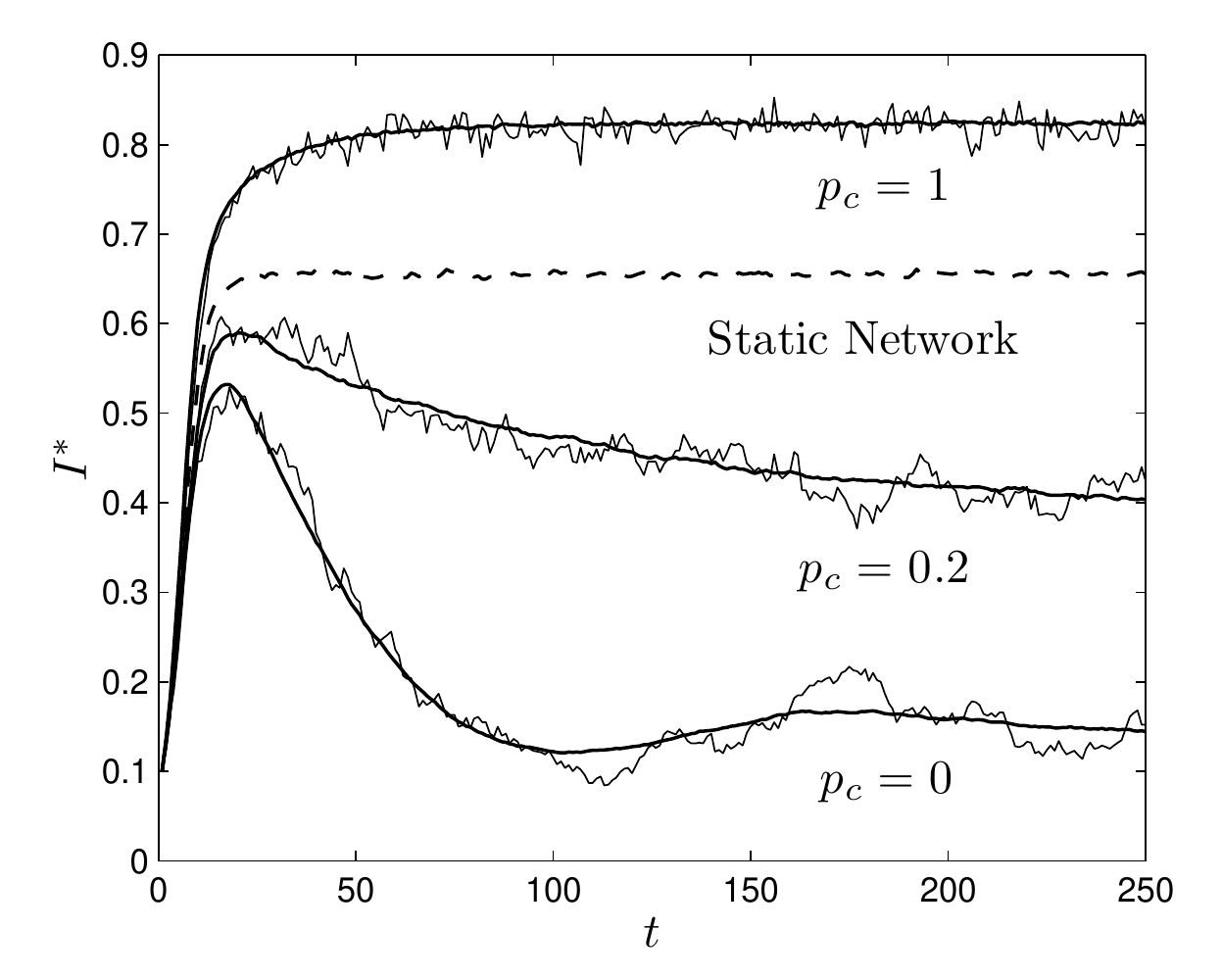}
\caption{Infected density ($I^{*}=I/N$) for SIS dynamics as a function of
time for different caretaker proportions $p_{c}$, where caretakers
do not improve recovery. \ER{} networks with adaptive rewiring were
used (solid lines), as well as a similar static network (no rewiring,
dashed line). Solid lines were obtained by averaging over 100 simulations,
so a single-simulation plot is overlaid in each adaptive scenario
for reference. The plot corresponds to $I_{0}=10^{2}$, $N=10^{3}$,
$p_{ER}=0.008,$ $\mu=0.05,$ $\gamma=0.037$, $\beta=0.15$, $\tau=0.18$.
\label{fig:SIS_time_course}}
\end{figure}

We first consider SIS dynamics. At each time step,
a randomly chosen node $i$ can transition from $S$ to $I$ with probability $p_{i}$,
or from $I$ to $S$ with probability $r_{i}$ as given above. To
study the effect of adaptive rewiring, we first consider a system without
the caretaker effect on the recovery rate, i.e.
$\beta_1=\beta_0$. Caretakers only increase their interaction with infected
individuals. 
We consider a network with weights initially distributed uniformly between
0 and 1. 
Results are shown in Fig.~\ref{fig:SIS_time_course}.
In the absence of caretaker links ($p_{c}=0)$, the equilibrium
endemic state $I^{*}=I_{t}/N$ is much lower than compared to the static
network (without rewiring). This is expected, as only regular (negative) interactions
exist that decrease in response to the epidemic. The total network
weight adapts to a smaller value, decreasing the endemic state. The
dynamics of the disease and adaptation of the network is visible in
the damped oscillation of the fraction of infecteds. 

However, as the
fraction of caretakers is increased, diseases can attain higher endemic
states than their static network counterparts. The caretaker dynamics
increases the interaction rate with infecteds, effectively yielding
a higher disease prevalence, which is expected.

\begin{figure}[t!]
\centering
\includegraphics[width=\columnwidth]{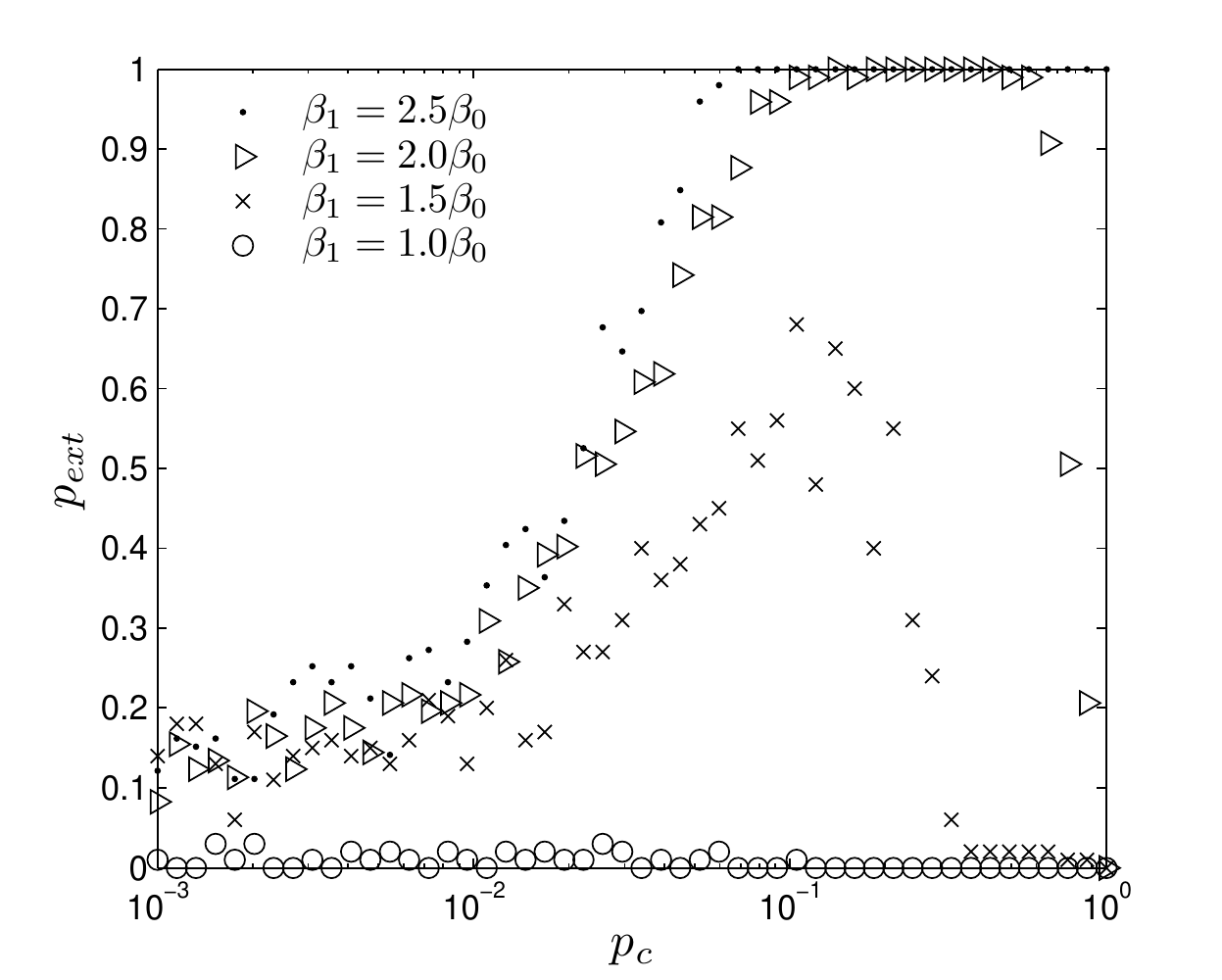}
\caption{Extinction probability $p_{ext}$ for SIS dynamics as a function of
caretaker proportion $p_{c}$ for various values of $\beta_{1}$ in an \ER{} network.
Note  a critical $p_{c}$ value at which extinction is
maximized. Approaching this value from the left yields a gradual increase
in extinction, while increasing $p_{c}$ past this critical value
causes a rapid decrease in $p_{ext}$.
The plot corresponds to $I_{0}=10^{2}$,
$N=10^{3}$, $\mu=0.05,$ $\gamma=0.037$, $\tau=0.18$, $\beta_{0}=0.35$,
$\sigma_{0}=\left<\sigma_{i}\right>\big|_{t=0}$, $p_{ER}=0.008$.
\label{fig:SIS_cross_sections}}
\end{figure}

\begin{figure*}[t!]
\centering
\includegraphics[scale=0.6]{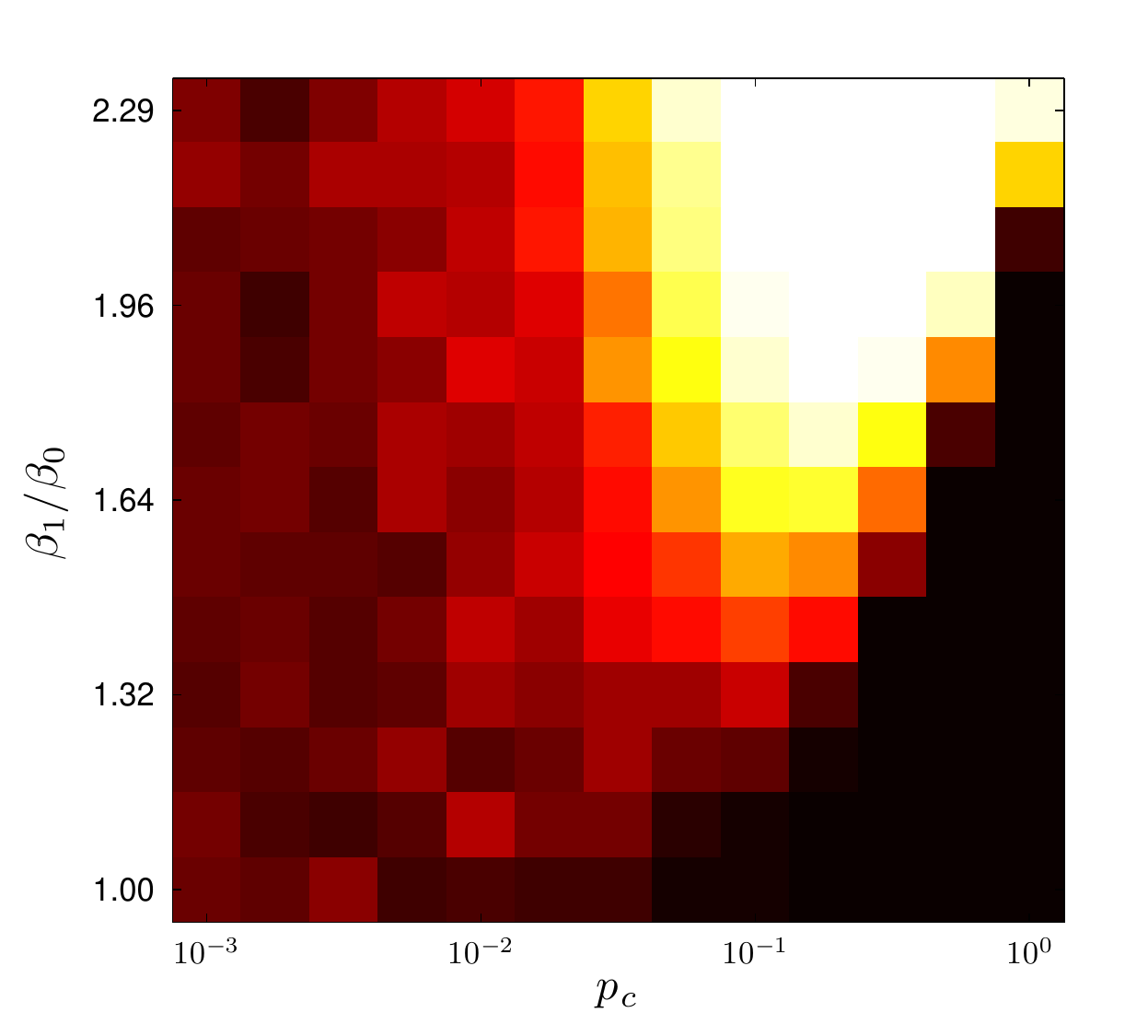} ~~~~~~~~~~~~
\includegraphics[scale=0.6]{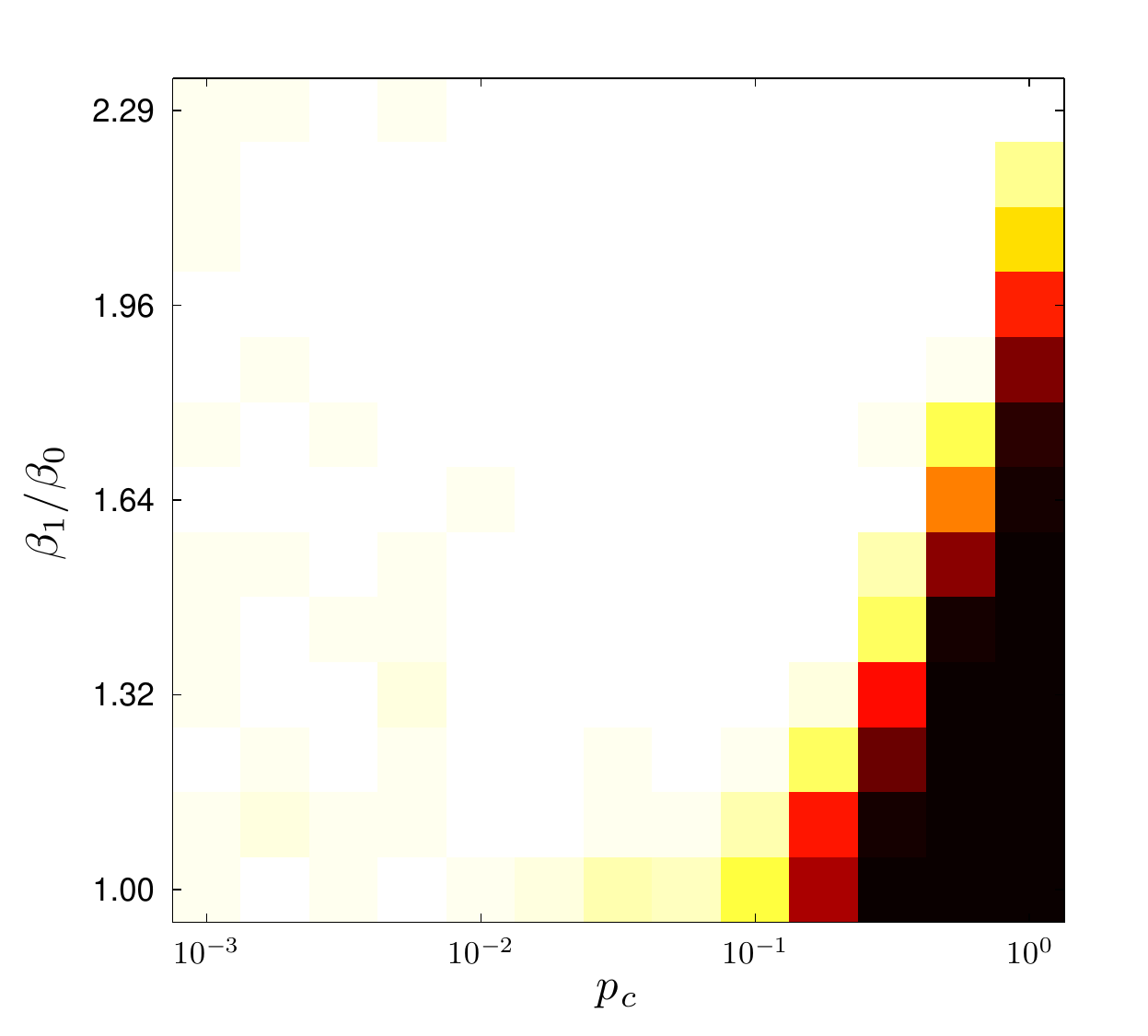}
\caption{Two-parameter phase diagrams showing extinction probability for SIS
dynamics as a function of maximum caretaker effectiveness $\beta_{1}$
and caretaker proportion $p_{c}$. \ER{} (left) and Scale-Free networks
(right) were considered. In the black regions, extinction probability
is 0 while extinction probability is 1 in the white regions. The plots
suggest that increasing the caretaker proportion past a critical value
yields a decreased extinction probability in both networks. On the
\ER{} network, $p_{c}\approx10^{-1}$ yields maximum disease extinction,
while extinction is most likely for $p_{c}\approx0$ on the Scale-Free
network. The plots correspond to $I_{0}=10^{2}$, $N=10^{3}$, $\mu=0.05,$
$\gamma=0.037$, $\tau=0.18$, $\beta_{0}=0.35$, $\sigma_{0}=\left<\sigma_{i}\right>\big|_{t=0}$
if $\left<\sigma_{i}\right>\big|_{t=0}>0$ otherwise $\beta_{i}=\beta_{0}$,
$p_{ER}=0.008$, (\ER{}) and a mean degree $k_0=2$ in the scale free network. 
\label{fig:Phase_Diagrams}}
\end{figure*}

The system that lacks a positive caretaker effect represents a somewhat
artificial limiting case. We therefore consider a positive \textit{caretaker
effect}: caretaker relationships lend higher recovery rates $\beta_1>\beta_0$ to
infected individuals, see Eq.~\eqref{eq:caretaker_effect}.  In particular, we
consider the effect of varying the maximum recovery rate $\beta_{1}$ and the
fraction of caretaker links $p_{c}$ on the extinction probability of the
disease. The results are depicted in Figs.~\ref{fig:SIS_cross_sections} and
\ref{fig:Phase_Diagrams}.  In general, increasing $\beta_{1}$ yields higher
extinction, since caretaker links are more effective at raising recovery rates.
One would then expect that increasing the caretaker proportion $p_{c}$ would
also yield higher extinction, as more relationships would cause increasing
recovery rates. However, this is not necessarily the case.  Raising the
caretaker proportion past some $\beta_{1}$-dependent critical value allows
diseases to persist. This critical value also serves as a threshold, as
increasing $p_{c}$ above this value rapidly decreases the extinction probability
to 0. This is illustrated in Fig.~\ref{fig:SIS_cross_sections}. Increasing
$p_{c}$ at first yields and increased $p_{ext}$ until a maximum is reached. A
further increase leads to a rapid decrease in extinction probability. For the
\ER{} network, the critical fraction of caretakers is approximately
$p_{c}\approx 10\%$.  For $p_{c}$ values above or below this, high extinction
probability is seen only for very high values of $\beta_{1}$. Note however, that
even for very small fractions of caretakers, a substantial increase in
extinction probability is observed. This suggests that, if the caretaker-effect
is taken into account, the best strategy to extinguish a disease is the
existence of a few effective caretaker relationships, that safely avoids the
negative effects that emerge beyond the critical concentration.  Note also that
for non-vanishing $p_{c}$, guaranteed extinction ($p_{ext}=1$) is observed only
for very high values of $\beta_{1}$. 

Note that these results were obtained for an \ER{} network.  In order to
investigate the interaction of network adaptation in combination with strong
network heterogeneity, we investigated the dynamics in a scale free topology.
The results are depicted in Fig.~\ref{fig:Phase_Diagrams}. In contrast to the
\ER{} system, we observe a high extinction of the disease for a wide range of
caretaker concentrations and recovery parameters $\beta_1$.  The disease is
endemic in the adaptive, scale free network only for small $\beta_1$ and large
$p_c$.  The implications of these results are interesting: In a scale free
adaptive network, regular links that decrease when connected to infected nodes
are sufficient to extinguish a disease, even in the presence of a considerable
fraction of caretaker links.  This strongly contrasts with the behavior observed
in static scale free networks, in which the existence of strongly connected hubs
generally facilitate the spread of a disease.  In the adaptive network, for
$p_{c}\ll1$, it is sufficient that the majority of nodes decrease their
interactions with the infected subpopulation.  In scale-free networks, hubs that
possess a large number of links will adaptively reduce the majority of their
regular weights, and thus their ability to serve as a gateway of the disease to
spread throughout the network.  In this regime, the effect of caretaker
relationships and their effect on recovery are benign.  Only when the fraction
of caretaker links reaches a large value such that also hubs become
predominantly caretakers, the situation changes, and the disease will evolve
into an endemic state.

\begin{figure*}[t!]
\centering 
\includegraphics[width=0.875\textwidth]{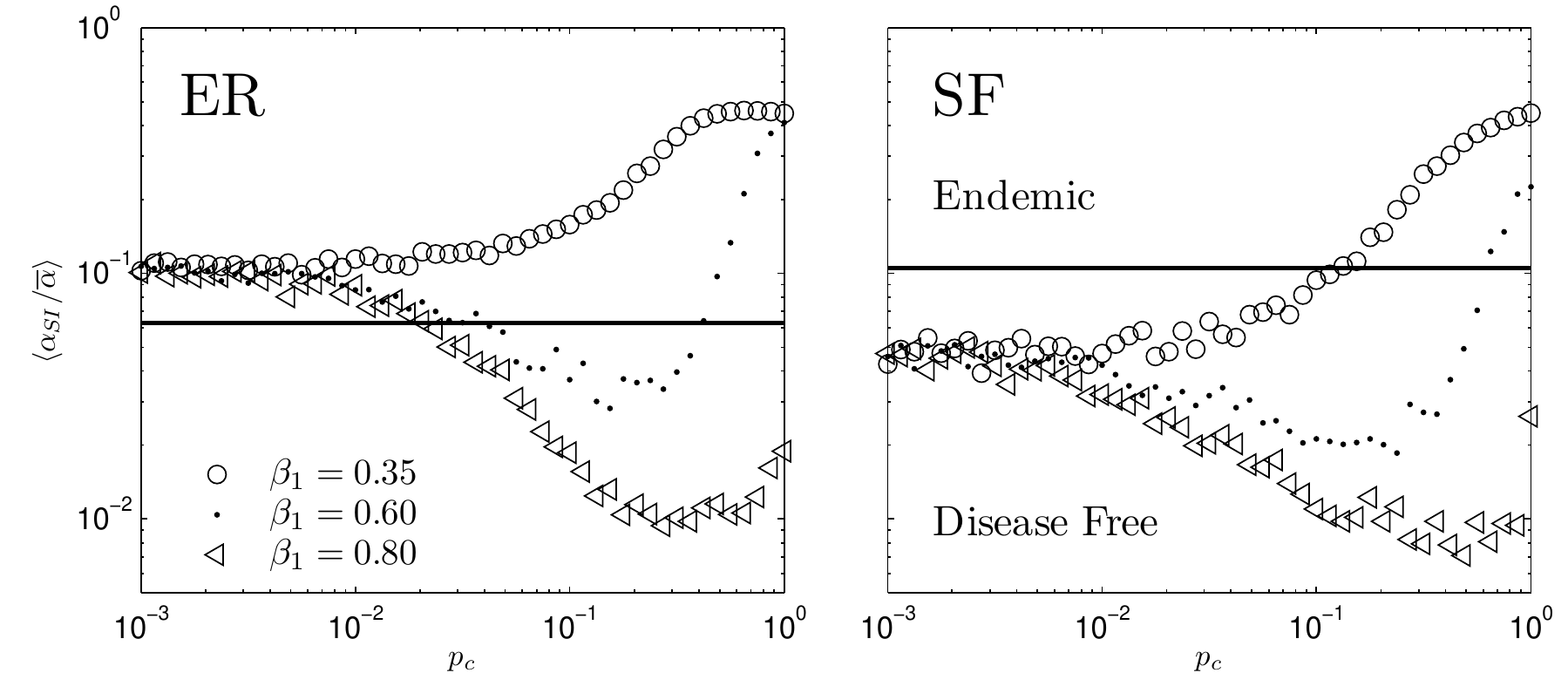}
\caption{Time-averaged SI contact fraction $\left<\alpha_{SI}\right>$
for SIS dynamics with different values of the caretaker proportion
$p_{c}$. Three $\beta_{1}$ values were chosen, 0.35 (circles), 0.60
(dots), 0.80 (arrows) to correspond with low, intermediate, and high
traces in the phase diagram of Fig.~\ref{fig:Phase_Diagrams}. An
\ER{} network was used (left), as well as a Scale-Free network (right).
The horizontal solid lines represent a critical value for $\left<\alpha_{SI}\right>$
above which the extinction probability vanishes and below which
the disease goes extinct. The plots correspond to $I_{0}=10^{2}$, $N=10^{3}$, $\mu=0.05,$
$\gamma=0.037$, $\tau=0.18$, $\beta_{0}=0.35$, $\sigma_{0}=\left<\sigma_{i}\right>\big|_{t=0}$
if $\left<\sigma_{i}\right>\big|_{t=0}>0$ otherwise $\beta_{i}=\beta_{0}$,
$p_{ER}=0.008$, (\ER{}) and mean degree $k_0=2$ (Scale-Free). \label{fig:Flux}
}
\end{figure*}

To explain these results, consider a susceptible node $i$ and its
total rate of interaction with infected neighbors: \[
\Phi_{SI}(i)=\sum_{j}w_{ij}x_{j}.\]
 The ratio of $SI$ interaction rates and total interaction rate $\alpha_{0}=\sum_{i<j}w_{ij}$
is given by \[
\alpha_{SI}=\frac{1}{\alpha_{0}}\sum_{i}\Phi_{SI}(i)(1-x_{i})\]
 Averaging this measure over the time-course of a disease gives us
a measure of the typical fraction of contacts due to SI interaction:
\[
    \left<\alpha_{SI}\right>=\frac{1}{T\alpha_{0}}\int_0^Tdt\,\left[\sum_{i,j}(1-x_{i})w_{ij}x_j\right]
\]
Now consider this time averaged $\left<\alpha_{SI}\right>$ as a function
of $p_{c}$ for various values of $\beta_{1}$, see Fig.~\ref{fig:Flux}.
For $\beta_1=\beta_0$ (i.e. no caretaker effect on recovery rates), the
rate of $SI$ interactions increase steadily as $p_c$ is increased, yielding
a more stable endemic state and high prevalence. When the caretaker effect
is taken into account, we observe an initial decrease of $SI$ interactions until
a critical value is reached below which the disease will go extinct, indicated by
the solid line. Increasing $p_c$ further can result in increasing $SI$ interactions
beyond this critical value, entering a regime in which a large fraction of caretaker links
results in a negative effect.
In the scale-free network, the qualitative behavior is similar. The crucial difference
is that typically, the adaptive process of regular links is sufficient to put the fraction of $SI$ links
below the critical value even in the absence of caretaker links.

\begin{figure}[t!]
    \centering
    \includegraphics[clip,width=\columnwidth]{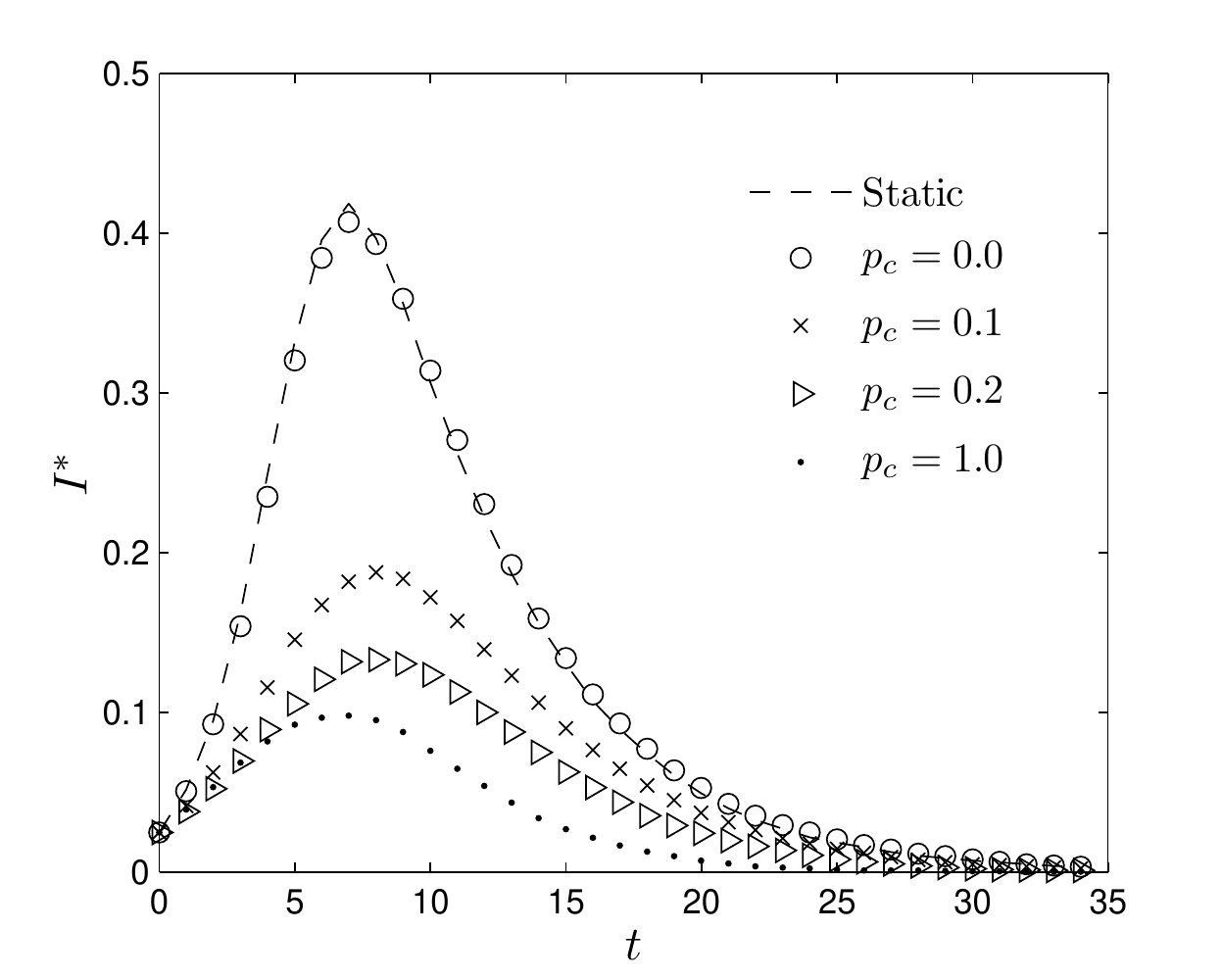}
    \caption{Infected density ($I^{*}=I/N$) for SIR dynamics as a function of
    time for different caretaker proportions $p_{c}$. \ER{} networks
    with adaptive rewiring were used, as well as a similar static network
    (no rewiring, dashed line). Increasing $p_{c}$ lowers the epidemic
    peak as well as the attack rate. Note also that the static network
    trace closely resembles the $p_{c}=0$ trace, showing that SIR diseases
    in this system are not significantly affected by dynamic link weights
    alone. The plots correspond to $I_{0}=25$, $N=10^{3}$, $\mu=0.05,$
    $\gamma=0.037$, $\tau=0.45$, $\beta_{0}=0.20$, $\sigma_{0}=\left<\sigma_{i}\right>\big|_{t=0}$
    if $\left<\sigma_{i}\right>\big|_{t=0}>0$ otherwise $\beta_{i}=\beta_{0}$,
    $n=2$, $p_{ER}=0.008$, (\ER{}). Scale-Free
    network results were similar. 
    \label{fig:SIR_time_course}}
\end{figure}

\begin{figure}[t!]
    \centering
\includegraphics[width=\columnwidth]{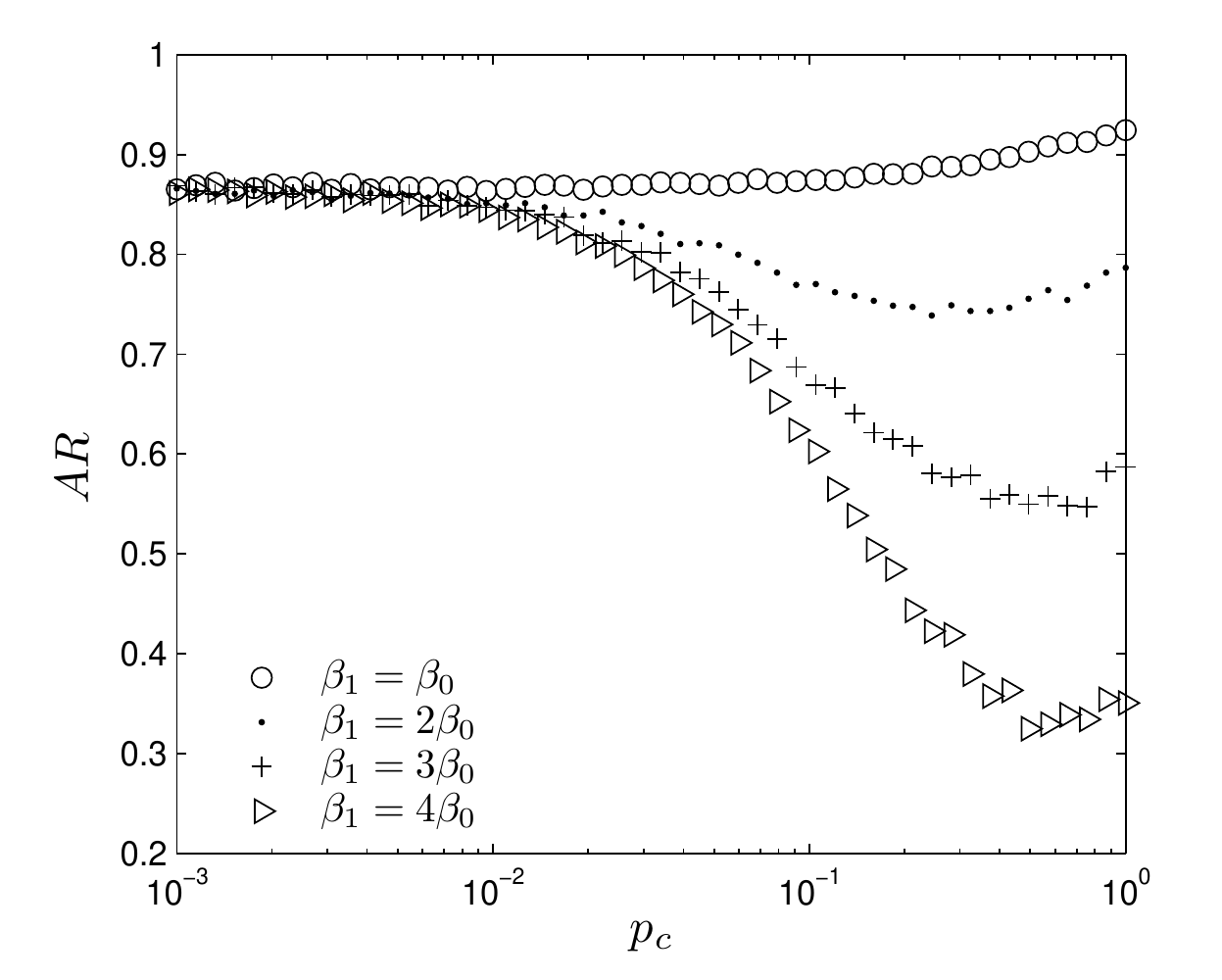}
\caption{Attack rate $AR$ as a function of $p_{c}$ for SIR dynamics with
various values of $\beta_{1}$ in an \ER{} network. For each $\beta_{1}>\beta_{0}$, the
attack rate is minimized for some value of $p_{c}$ between $10^{-1}$
and $10^{0}$. As $\beta_{1}$ increases, this minimum point shifts
subtly to the right. This shows that the more effective caretakers
are at healing, the more caretaker relationships the system can permit
before they have a negative impact on the attack rate. The plots correspond
to $I_{0}=25$, $N=10^{3}$, $\mu=0.05,$ $\gamma=0.037$, $\tau=0.25$,
$\beta_{0}=0.20$, $\sigma_{0}=\left<\sigma_{i}\right>\big|_{t=0}$
if $\left<\sigma_{i}\right>\big|_{t=0}>0$ otherwise $\beta_{i}=\beta_{0}$,
$n=2$, $p_{ER}=0.008$.
\label{fig:AR_cross_sections}}
\end{figure}

Next we turn out attention to the effect of caretaker adaptive networks on systems
that are better described by SIR dynamics.
Here individuals (nodes) exist in one of three states, susceptible
(S), infected (I) or recovered (R). Individuals can transition from
$S$ to $I$ with probability $p_{i}$ and from $I$ to $R$ with
probability $r_{i}$, as given above in Eqs.~\eqref{eq:infection_probability}
and \eqref{eq:caretaker_effect}. The state $R$ is absorbing, so
once all infected nodes in a population recover, the disease dies
out (see Fig.~\ref{fig:SIR_time_course}). 
In order to investigate the impact of caretaker dynamics and an SIR scenario,
we focus on the \textit{attack rate (ratio)} and the \textit{epidemic
peak}. The attack ratio (AR) is simply the fraction of the population
which contracts the infection at some point during the epidemic. Since
every infected node eventually enters the recovered class, this is
equivalent to the fraction of recovered nodes at the end of the epidemic:
\[
AR=\frac{R_{\infty}}{N}
\]
The epidemic peak (EP) is the maximum infected fraction attained in the
population over the course of the epidemic.
Figure \ref{fig:AR_cross_sections} depicts the attack ratio as a function of
$p_c$ for various values of the recovery rate parameter $\beta_1$.
Interestingly, without a caretaker effect ($\beta_1=\beta_0$) the increase
in attack ration is not substantial as $p_c$ is increased. For $\beta_1>\beta_0$,
we observe a decrease in attack ratio even for small fractions of caretaker links.
The minimum attack ratio is attained only in a regime where most links are caretaker links.

\begin{figure*}[t!]
\centering
\includegraphics[clip,scale=0.6]{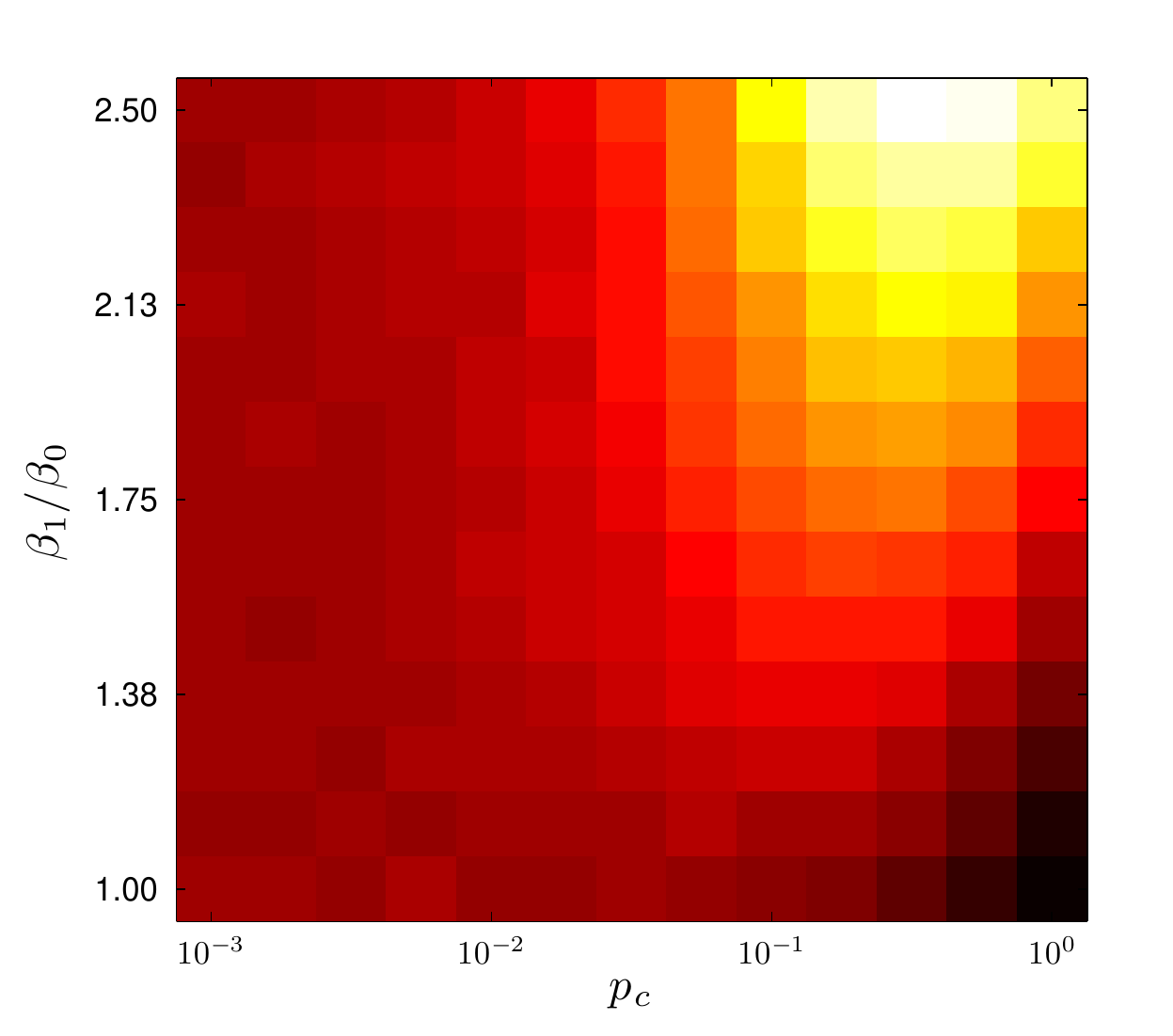}  ~~~~~~~~~~~~ 
\includegraphics[clip,scale=0.6]{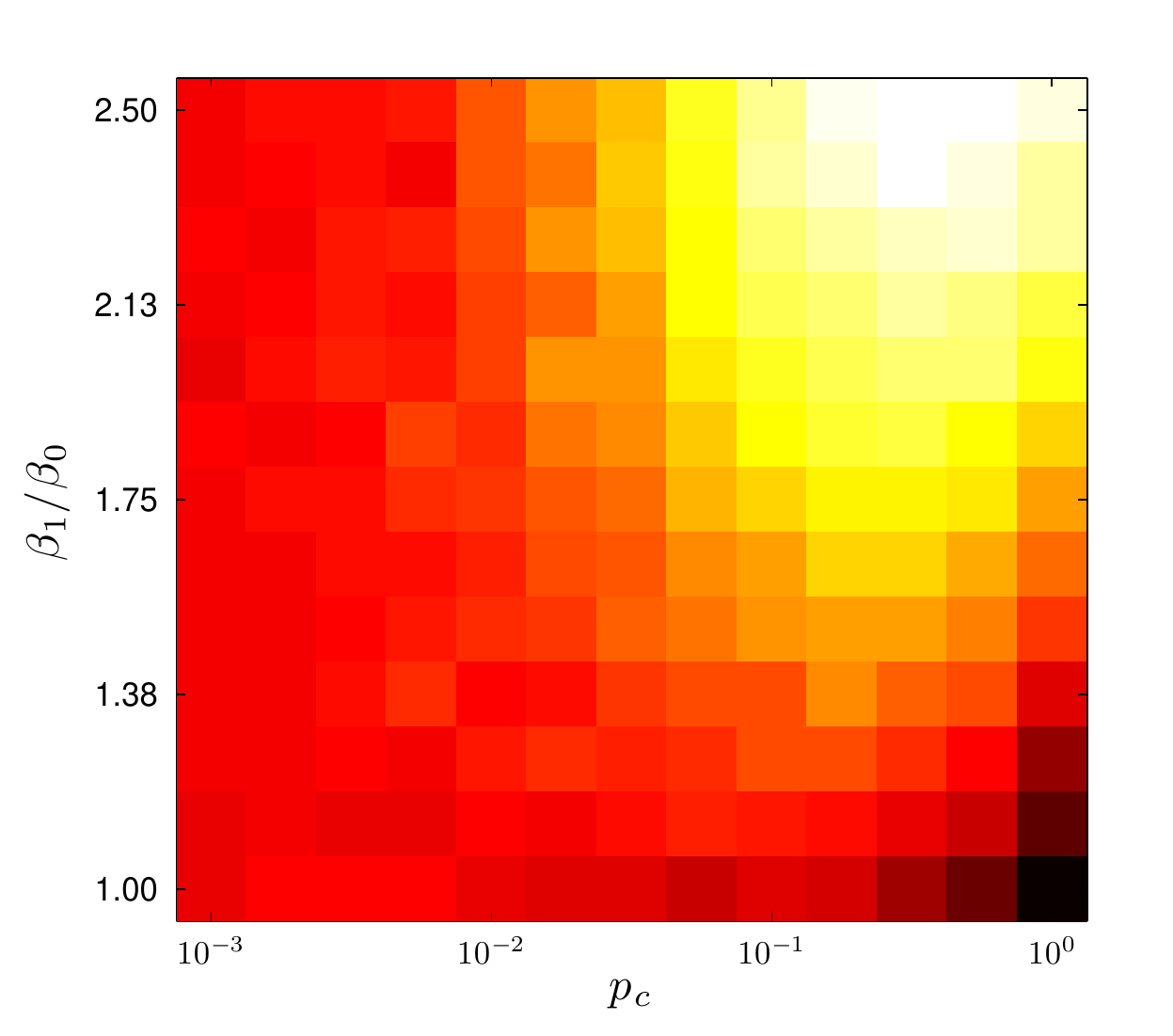}
\caption{Two-parameter phase diagrams showing the dependence of attack rate
in SIR dynamics on maximum caretaker effectiveness $\beta_{1}$ (normalized
by the baseline-recovery probability $\beta_{0}$) and caretaker proportion
$p_{c}$. \ER{} (left) and Scale-Free (right) networks were considered.
Attack rate approaches zero in the white regions, while it approaches
$1$ in the black regions. Note that increasing $p_{c}$ yields lower
attack rates for $p_{c}<0.2$, but increasing past this critical value
yields increasing attack rates. There is a critical value $p_{c}\approx0.2$
at which attack rate is minimized for most values of $\beta_{1}$.
Furthermore, this effect is seen in both ER and SF networks, though
attack rates are lower overall on the SF network. The plots correspond
to $I_{0}=25$, $N=10^{3}$, $\mu=0.05,$ $\gamma=0.037$, $\tau=0.25$,
$\beta_{0}=0.20$, $\sigma_{0}=\left<\sigma_{i}\right>\big|_{t=0}$
if $\left<\sigma_{i}\right>\big|_{t=0}>0$ otherwise $\beta_{i}=\beta_{0}$,
$n=2$, $p_{ER}=0.008$, (\ER{}) and mean degree $k_0=2$ (Scale-Free). 
\label{fig:attack_rate_phase}}
\end{figure*}

\begin{figure*}[t!]
\centering
\includegraphics[clip,scale=0.6]{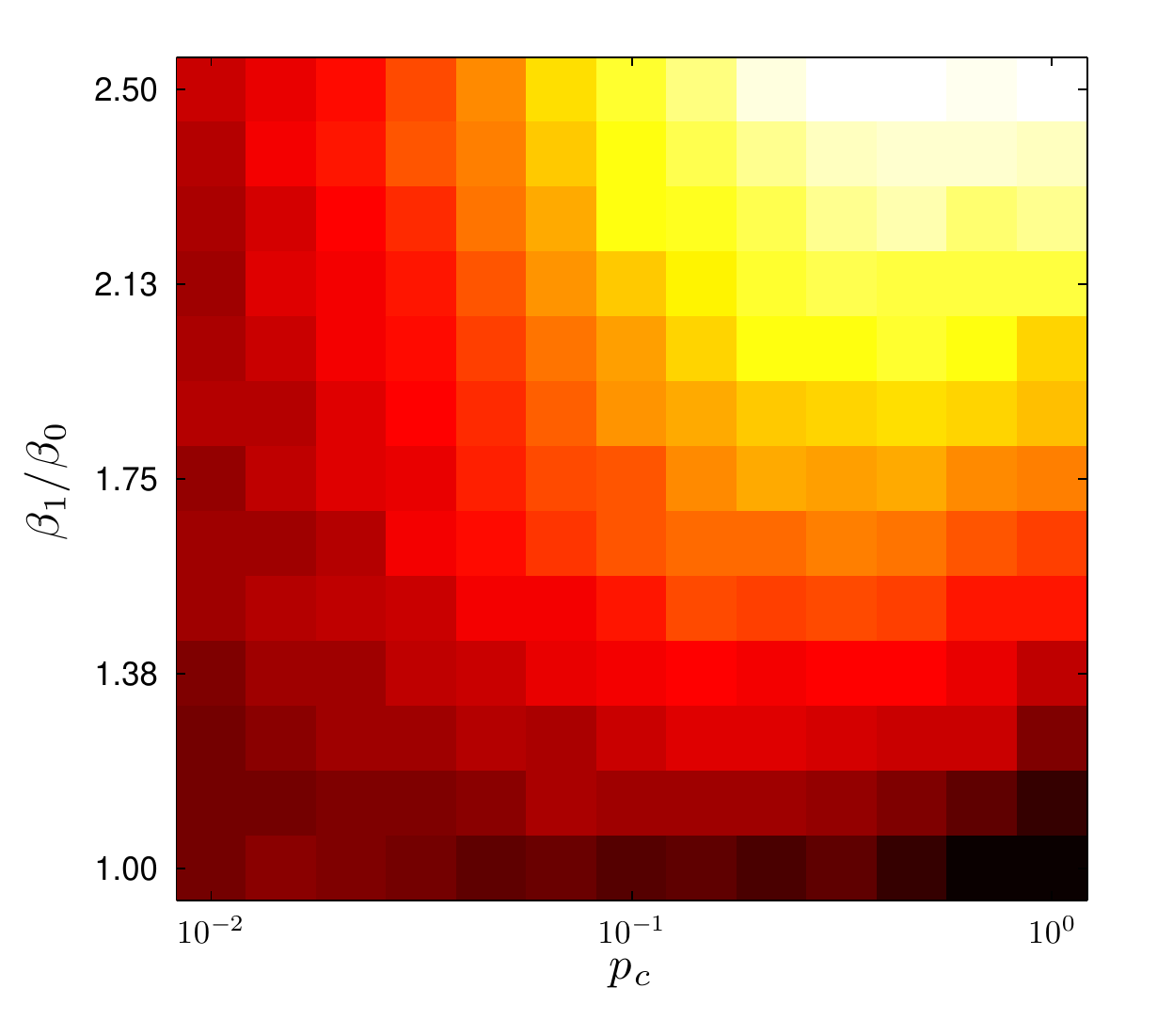}  ~~~~~~~~~~~~ 
\includegraphics[clip,scale=0.6]{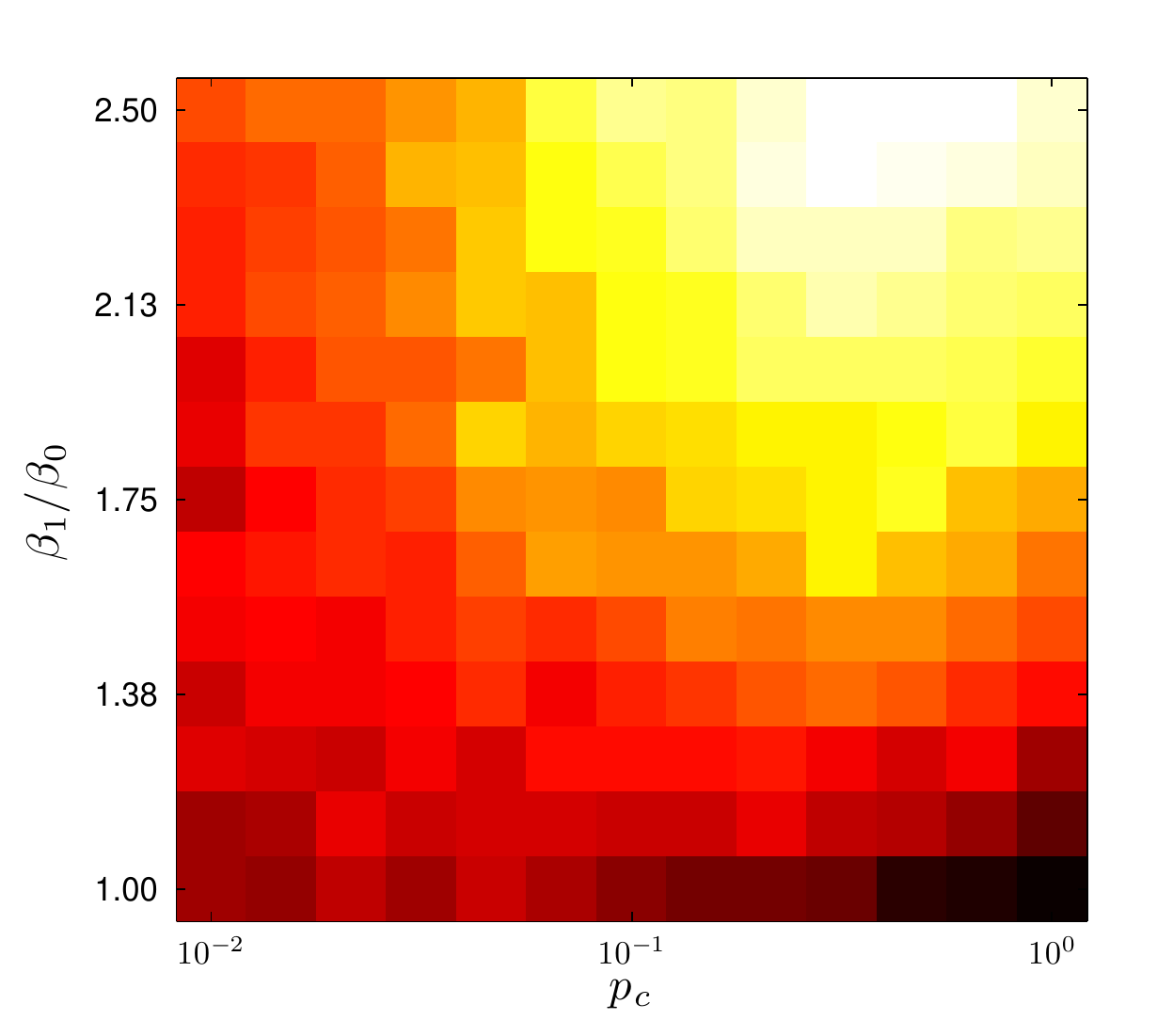}
\caption{Two-parameter phase diagrams showing the dependence of the epidemic
peak $(EP)$ in SIR dynamics on the maximum caretaker effectiveness
$\beta_{1}$ (normalized by the baseline-recovery probability $\beta_{0}$)
and caretaker proportion $p_{c}$. \ER{} (a) and Scale-Free (b)
networks were considered. The epidemic peak approaches zero in the
white regions, while it approaches 1 in the black regions. Note the
similarities to the attack rate diagram in Fig.~\ref{fig:attack_rate_phase}.
The epidemic peak is minimized for $p_{c}\approx0.2$ for most values
of $\beta_{1}$, but for $p_{c}<0.2$ or $p_{c}>0.2$, the attack
rate is greater for a given value of $\beta_{1}$. The plots correspond
to $I_{0}=25$, $N=10^{3}$, $\mu=0.05,$ $\gamma=0.037$, $\tau=0.40$,
$\beta_{0}=0.20$, $\sigma_{0}=\left<\sigma_{i}\right>\big|_{t=0}$
if $\left<\sigma_{i}\right>\big|_{t=0}>0$ otherwise $\beta_{i}=\beta_{0}$,
$n=2$, $p_{ER}=0.008$, (\ER{}) and mean degree $k_0=2$ (Scale-Free). 
\label{fig:epidemic_peak}}
\end{figure*}

Figure \ref{fig:attack_rate_phase} depicts the attack ratio as a function
of both system parameters $\beta_1$ and $p_c$ and compares the behavior
in both network architectures, \ER{} and \BA{}. In contrast with the SIR system,
network topology does not substantially change the dynamics, both networks
exhibit a similar attack ratio as a function of $\beta_1$ and $p_c$.
For fixed $\beta_1$ increasing $p_c$ first decreases the attack ratio until
a minimum is attained. Increasing $p_c$ further increases the attack ratio again.
A consistent effect is observed in the response of the epidemic peak to changes
in $\beta_1$ and $p_c$, see Fig.~\ref{fig:epidemic_peak}.

The dynamics seen above for the attack rate are mirrored in the epidemic
peak $EP$ as well (Fig.~\ref{fig:epidemic_peak}), which decreases
as caretaker effectiveness (represented by $\beta_{1}$) increases.
There is again a critical relationship with $p_{c}$, as values of
$p_{c}\approx0.2$ tend to minimize the epidemic peak for $\beta_{1}>\beta_{0}$.
Again though, for $\beta_{1}=\beta_{0}$, increasing $p_{c}$ yields
a monotonic increase in $EP$.

\section{Conclusions}

Individual response can have a great impact on the dynamics of spreading
diseases on complex networks. In particular, if one uses an avoidance strategy
whereby all individuals simply avoid infecteds, the endemic state of an SIS
disease can be drastically reduced.  On the other hand, allowing individuals
(caretakers) to become closer to infecteds is a calculated risk. If the
caretakers are not effective healers (such as non-physician parents and
children), then the severity of the disease generally increases. But if the
caretakers are effective healers (consider doctor/patient relationships, for
example), then the outcome of the disease is generally improved even by a small
number of them.  If too many caretakers are introduced, though, their healing
benefit is overridden by their increased exposure, yielding a worse outcome than
if the population had simply not reacted.

These findings have a number of implications in public health. For one, in a
large-scale epidemic there certainly exists a critical fraction of doctors and
aid workers in the population. If there are too few or too many, they can
actually \textit{increase} the total number of individuals infected over the
course of the disease. In such cases, it would actually be more beneficial to
employ an avoidance strategy whereby all individuals, including doctors and aid
workers, simply avoided infected individuals. In the particular case of $SIS$
endemic diseases, we have seen that the critical caretaker proportion is
actually $p_{c}=0$ on Scale-Free networks. This suggests that networks that
exhibit a strong variability in interaction statistics and at the same time are
adaptive, are less susceptible to the risk of endemic diseases, and the natural
instinct to avoid infection is more effective in eliminating a disease than the
positive effects that caretakers may have.


\end{document}